\documentclass[twocolumn,english,aps,prl,showpacs,superscriptaddress]{revtex4-1}
\usepackage[latin9]{inputenc}
\usepackage{amsmath}
\usepackage{amssymb}
\usepackage{wasysym}
\usepackage{graphicx}
\usepackage{esint}

\makeatletter

\@ifundefined{textcolor}{}
{%
 \definecolor{BLACK}{gray}{0}
 \definecolor{WHITE}{gray}{1}
 \definecolor{RED}{rgb}{1,0,0}
 \definecolor{GREEN}{rgb}{0,1,0}
 \definecolor{BLUE}{rgb}{0,0,1}
 \definecolor{CYAN}{cmyk}{1,0,0,0}
 \definecolor{MAGENTA}{cmyk}{0,1,0,0}
 \definecolor{YELLOW}{cmyk}{0,0,1,0}
}
\newenvironment{lyxlist}[1]
{\begin{list}{}
{\settowidth{\labelwidth}{#1}
 \setlength{\leftmargin}{\labelwidth}
 \addtolength{\leftmargin}{\labelsep}
 }}
{\end{list}}

\usepackage{babel}

\usepackage{babel}

\makeatother

\usepackage{babel}
\begin{document}

\title{High-fidelity and robust two-qubit gates for quantum-dot spin qubits
in silicon}

\author{Chia-Hsien Huang}

\affiliation{Department of Physics and Center for Theoretical Physics, National
Taiwan University, Taipei 10617, Taiwan}

\affiliation{Center for Quantum Science and Engineering, National Taiwan University,
Taipei 10617, Taiwan}

\author{C. H. Yang}

\affiliation{Centre for Quantum Computation and Communication Technology, School
of Electrical Engineering and Telecommunications, The University of
New South Wales, Sydney, New South Wales 2052, Australia}

\author{Chien-Chang Chen}

\affiliation{Department of Physics and Center for Theoretical Physics, National
Taiwan University, Taipei 10617, Taiwan}

\affiliation{Center for Quantum Science and Engineering, National Taiwan University,
Taipei 10617, Taiwan}

\author{A. S. Dzurak}

\affiliation{Centre for Quantum Computation and Communication Technology, School
of Electrical Engineering and Telecommunications, The University of
New South Wales, Sydney, New South Wales 2052, Australia}

\author{Hsi-Sheng Goan}

\email{goan@phys.ntu.edu.tw}

\selectlanguage{english}%

\affiliation{Department of Physics and Center for Theoretical Physics, National
Taiwan University, Taipei 10617, Taiwan}

\affiliation{Center for Quantum Science and Engineering, National Taiwan University,
Taipei 10617, Taiwan}

\date{\today}
\begin{abstract}
A two-qubit controlled-NOT (CNOT) gate, realized by a controlled-phase
(C-phase) gate combined with single-qubit gates, has been experimentally
implemented recently for quantum-dot spin qubits in isotopically enriched
silicon, a promising solid-state system for practical quantum computation.
In the experiments, the single-qubit gates have been demonstrated
with fault-tolerant control-fidelity, but the infidelity of the two-qubit
C-phase gate is, primarily due to the electrical noise, still higher
than the required error threshold for fault-tolerant quantum computation
(FTQC). Here, by taking the realistic system parameters and the experimental
constraints on the control pulses into account, we construct experimentally
realizable high-fidelity CNOT gates robust against electrical noise
with the experimentally measured $1/f^{1.01}$ noise spectrum and
also against the uncertainty in the interdot tunnel coupling amplitude.
Our optimal CNOT gate has about two orders of magnitude improvement
in gate infidelity over the ideal C-phase gate constructed without
considering any noise effect. Furthermore, within the same control
framework, high-fidelity and robust single-qubit gates can also be
constructed, paving the way for large-scale FTQC. 
\end{abstract}


\maketitle
\global\long\def\figurename{FIG.}
 \makeatletter \global\long\def\thefigure{\@arabic\c@figure}
 \makeatother

Electron spin qubits in semiconductor quantum dots \cite{LossDiVincenzo1998}
are promising solid-state systems to realize quantum computation.
Significant progresses of quantum-dot spin qubits for quantum information
processing have been made with %
\mbox{%
III%
}-%
\mbox{%
V%
} semiconductors such as GaAs \cite{PettaJohnsonTaylorEtAl2005,KoppensBuizertTielrooijEtAl2006,KoppensNowackVandersypen2008,ReillyTaylorPettaEtAl2008,FolettiBluhmMahaluEtAl2009,BluhmFolettiMahaluEtAl2010,BarthelMedfordMarcusEtAl2010,BluhmFolettiNederEtAl2011,NowackShafieiLaforestEtAl2011,BrunnerShinObataEtAl2011,ShulmanDialHarveyEtAl2012,MedfordJ.BeilJ.TaylorJ.EtAl2013,NicholOronaHarveyEtAl2017},
but the coherence time of the qubits is limited by the strong dephasing
from the environment nuclear spins \cite{ChekhovichMakhoninTartakovskiiEtAl2013}.
On the other hand, the coherence time is substantially improved by
using a Si-based host substrate \cite{MauneBorselliHuangEtAl2012,KawakamiE.ScarlinoP.WardD.EtAl2014,VeldhorstM.HwangJ.C.YangC.EtAl2014,VeldhorstYangHwangEtAl2015,EngLaddSmithEtAl2015,TakedaKamiokaOtsukaEtAl2016,Kawakami18102016,YonedaTakedaOtsukaEtAl2018,ZajacSigillitoRussEtAl2018,WatsonPhilipsKawakamiEtAl2018}.
For qubits in isotopically enriched $^{28}{\rm Si}$, the decoherence
(dephasing) time $T_{2}^{\star}$ can be further extended to $120\mu s$
\cite{VeldhorstM.HwangJ.C.YangC.EtAl2014,VeldhorstYangHwangEtAl2015}.
So far, the single-qubit gates for silicon-based quantum-dot spin-qubit
systems have been demonstrated with fault-tolerant control-fidelity
\cite{VeldhorstM.HwangJ.C.YangC.EtAl2014,VeldhorstYangHwangEtAl2015,TakedaKamiokaOtsukaEtAl2016,YonedaTakedaOtsukaEtAl2018,ZajacSigillitoRussEtAl2018}.
The two-qubit gates have also been realized \cite{VeldhorstYangHwangEtAl2015,ZajacSigillitoRussEtAl2018,WatsonPhilipsKawakamiEtAl2018},
but their fidelities have not yet reached the criterion for fault-tolerant
quantum computation (FTQC), primarily due to the noise of the electrical
voltage control used to realize the two-qubit gate. Some theoretical
pulse-design schemes to improve fidelity for two-qubit gates have
been proposed \cite{RussZajacSigillitoEtAl2018,gungordu2018pulse}.

The goal of this paper is to construct experimentally realizable robust
two-qubit gates for quantum-dot spin qubits in isotopically enriched
silicon with fidelity enabling large-scale FTQC. To this end, we apply
a robust control method \cite{HuangGoan2017} to suppress the electrical
noise with the experimentally measured $1/f^{1.01}$ noise spectrum
\cite{YonedaTakedaOtsukaEtAl2018} using the realistic system parameters
\cite{VeldhorstYangHwangEtAl2015}. The experimental constraint on
the maximum ac magnetic field strength due to the power limitation
through the on-chip electron spin resonance (ESR) line and the filtering
effects on the control pulses due to the finite bandwidth of waveform
generators are also accounted for. Due to the expected long coherence
time in isotopically enriched silicon system, we do not consider the
single-qubit decoherence (dephasing) noise in our calculations as
it does not affect appreciably the performance of the gates we construct
\cite{supplementary}. Instead of decomposing a CNOT gate into a C-phase
gate and several single-qubit gates in series as in the experiment
\cite{VeldhorstYangHwangEtAl2015}, we can construct single smooth
pulses for the high-fidelity CNOT gates directly to reduce the gate
operation time and the accumulated gate errors from the decomposed
gates. Compared with the ideal C-phase gate constructed without considering
any noise, our optimal CNOT gate can improve the fidelity loss from
the electrical noise by near two orders of magnitude (fidelity=99.997\%)
and enlarge the robust window against the uncertainty of the system
parameter by about 10 times. Besides, our smooth pulses with zero
strength and zero derivative at the initial and final gate operation
times can avoid the fidelity-loss due to the rise time and fall time
issues between the pulse-pulse connections of adjacent gate operations.
We also investigate other possible $1/f^{\alpha}$ noise spectra with
$0.7\leq\alpha<1.01$, and demonstrate that for the case of $\alpha=0.7$,
the infidelity of the high-fidelity CNOT gates by the same control
method \cite{HuangGoan2017} under the same experimental constraints
can still have one order of magnitude improvement over the ideal C-phase
gate.

In our scheme, the detuning energy is kept to a constant value when
operating a sequence of single-qubit and two-qubit gates. In contrast,
in the experiment \cite{VeldhorstYangHwangEtAl2015}, a single-qubit
gate is realized by tuning down the detuning energy (or relative alignment
potential of the two dots) to a small constant value as compared to
the on-site double-occupancy Coulomb energy to decouple the two-qubit
coupling; inversely, a two-qubit gate is realized by tuning up the
detuning energy to a large constant value to increase the coupling
between the two qubits. However, when operating a sequence of single-qubit
gates and two-qubit gates, the rise and fall times of the detuning
energy between two-qubit gate and single-qubit gates would cause gate
errors. Besides, changing detuning energy accompanies stark shifts
on the quantum-dot qubits, which may result in additional gate errors
if the calibration is not precise. Therefore, to prevent the fidelity
degradation from tuning the detuning energy up and down, we propose
to operate a sequence of single-qubit and two-qubit gates with the
detuning energy fixed.

In the following, we first introduce the ideal system of the quantum-dot
spin qubits in isotopically enriched silicon \cite{VeldhorstYangHwangEtAl2015},
then analyze the factors that degrade the gate fidelity in a realistic
system, after that briefly introduce the robust control method \cite{HuangGoan2017},
and finally demonstrate the performance of high-fidelity and robust
CNOT and single-qubit gates in the same control framework, i.e., with
detuning energy fixed and ac magnetic field as the control field.

For the quantum-dot electron spin qubits in isotopically enriched
silicon, the ideal two-qubit Hamiltonian written in the basis states
of ($\left|\textrm{dot2},\textrm{dot1}\right\rangle =$) $\left|\uparrow,\uparrow\right\rangle $,
$\left|\uparrow,\downarrow\right\rangle $, $\left|\downarrow,\uparrow\right\rangle $,
$\left|\downarrow,\downarrow\right\rangle $ and $\left|0,2\right\rangle $
can be expressed as\setlength{\arraycolsep}{0.4pt} 
\begin{align}
 & \mathcal{H}_{I}(t)/h=\nonumber \\
 & \left(\begin{array}{ccccc}
\overline{E}_{Z} & \frac{1}{2}E_{X}(t) & \frac{(1+\eta)}{2}E_{X}(t) & 0 & 0\\
\frac{1}{2}E_{X}(t) & \frac{1}{2}\delta E_{Z} & 0 & \frac{(1+\eta)}{2}E_{X}(t) & t_{0}\\
\frac{(1+\eta)}{2}E_{X}(t) & 0 & -\frac{1}{2}\delta E_{Z} & \frac{1}{2}E_{X}(t) & -t_{0}\\
0 & \frac{(1+\eta)}{2}E_{X}(t) & \frac{1}{2}E_{X}(t) & -\overline{E}_{Z} & 0\\
0 & t_{0} & -t_{0} & 0 & U-\epsilon(t)
\end{array}\right),\label{eq:qd_ideal_hamiltonian}
\end{align}
where $h$ is the Plank constant, $\overline{E}_{Z}=(E_{Z_{1}}+E_{Z_{2}})/2$
is the average frequency and $\delta E_{Z}=(E_{Z_{2}}-E_{Z_{1}})$
is the frequency difference of Zeeman splitting in the z-direction
for dot1 and dot2, $E_{Z_{1}}$ and $E_{Z_{2}}$, respectively, $t_{0}$
is the interdot tunnel coupling and $hU$ is the on-site Coulomb energy,
and $h\epsilon$ is the detuning energy or relative alignment of the
potential of the two dots. In principle, Zeeman splitting frequency
in the $x$-direction for dot1 and dot2 can be different and denoted
as $E_{X}(t)$ and $(1+\eta)E_{X}(t)$, respectively, where $E_{X}(t)=g\mu_{B}B_{X}(t)/h$
. Here $\eta$ is the $x$-direction $g$ factor difference fraction
between two dots, and the corresponding value for the $z$-direction
is $\sim0.001$ in the experiment \cite{VeldhorstYangHwangEtAl2015}.
Without losing generality, we choose $\eta=0$ to demonstrate the
gate performance here. We have examined the controllability of $\mathcal{H}_{I}(t)$
of Eq. (\ref{eq:qd_ideal_hamiltonian}) for CNOT gates and single-qubit
gates, and the same level of performance as $\eta=0$ case can be
achieved for $\left|\eta\right|\leq0.1$ cases. We control ac magnetic
field $B_{X}(t)=\Omega_{X}(t)\cos(\overline{E}_{Z}2\pi t)+\Omega_{Y}(t)\cos(\overline{E}_{Z}2\pi t+\dfrac{\pi}{2})$
via an on-chip ESR line with amplitudes $\Omega_{X}(t)$ and $\Omega_{Y}(t)$
to operate quantum gates. In the experiment \cite{VeldhorstYangHwangEtAl2015},
the C-phase gate is realized by tuning the detuning energy $\epsilon$
to a constant value {[}with the ac magnetic field $B_{X}(t)$ off{]}
to accumulate the time-integrated phase shift via the effective detuning
frequency $\nu_{\uparrow\downarrow,(\downarrow\uparrow)}$. There,
the system parameters $\overline{E}_{Z}=39.16{\rm GHz}$, $\delta E_{Z}=-40{\rm MHz}$,
and $t_{0}=900{\rm MHz}$. We will use also these realistic system
parameters to construct high-fidelity and robust CNOT gates and single-qubit
gates demonstrated later.

We define the ideal gate infidelity as $J_{1}\equiv1-|{\rm Tr}[U_{T}^{\dagger}U_{I,4\times4}(t_{f})]|^{2}/16$,
where ${\rm Tr}$ denotes a trace over the 2-qubit system state space,
$U_{T}$ is the two-qubit target gate, and $U_{I,4\times4}(t_{f})$
is the projected propagator in the subspace spanned by the two-qubit
computational basis states $\left\{ \left|\uparrow,\uparrow\right\rangle ,\left|\uparrow,\downarrow\right\rangle ,\left|\downarrow,\uparrow\right\rangle ,\left|\downarrow,\downarrow\right\rangle \right\} $
obtained from the ideal system propagator $U_{I}(t_{f})=\mathcal{T}_{+}\exp\left[-(i/\hbar)\int_{0}^{t_{f}}\mathcal{H}_{I}(t')dt'\right]$
at the final gate operation time $t_{f}$, where $\mathcal{T}_{+}$
is the time-ordering operator. In the definition of the ideal gate
infidelity $J_{1}$ , the leakage error, i.e., the state probability
remains in the $\left|0,2\right\rangle $ subspace, is also accounted
for.

However, in a realistic system, there exist many factors degrading
the gate fidelity such as the electrical noise $\beta_{U-\epsilon}(t)$,
the uncertainty $\alpha_{t_{0}}$ in tunnel coupling $t_{0}$, and
the filtering effects on the control pulses due to the finite bandwidth
of waveform generators. So, a realistic Hamiltonian taking these factors
into account becomes\setlength{\arraycolsep}{0.4pt} 
\begin{align}
 & \mathcal{H}(t)/h=\nonumber \\
 & \left(\begin{array}{ccccc}
\overline{E}_{Z} & \frac{1}{2}E_{X}^{{\rm filt}}(t) & \frac{1}{2}E_{X}^{{\rm filt}}(t) & 0 & 0\\
\frac{1}{2}E_{X}^{{\rm filt}}(t) & \frac{1}{2}\delta E_{Z} & 0 & \frac{1}{2}E_{X}^{{\rm filt}}(t) & (t_{0}+\alpha_{t_{0}})\\
\frac{1}{2}E_{X}^{{\rm filt}}(t) & 0 & -\frac{1}{2}\delta E_{Z} & \frac{1}{2}E_{X}^{{\rm filt}}(t) & -(t_{0}+\alpha_{t_{0}})\\
0 & \frac{1}{2}E_{X}^{{\rm filt}}(t) & \frac{1}{2}E_{X}^{{\rm filt}}(t) & -\overline{E}_{Z} & 0\\
0 & (t_{0}+\alpha_{t_{0}}) & -(t_{0}+\alpha_{t_{0}}) & 0 & U-\epsilon+\beta_{U-\epsilon}(t)
\end{array}\right),\label{eq:qd_real_hamiltonian}
\end{align}
where $E_{X}^{{\rm filt}}(t)=(g\mu_{B}/h)[\Omega_{X}^{{\rm filt}}(t)\cos(\overline{E}_{Z}2\pi t)$$+\Omega_{Y}^{{\rm filt}}(t)\cos(\overline{E}_{Z}2\pi t+\dfrac{\pi}{2})]$
with $\Omega_{X}^{{\rm filt}}(t)$ and $\Omega_{Y}^{{\rm filt}}(t)$
being the actual output control pulses on the qubits with the filtering
effects accounted for. We assume the electrical noise $\beta_{U-\epsilon}(t)$
is accompanied by the electrical control of the detuning energy $\epsilon$
and appears in the same location of $\epsilon$ in the Hamiltonian.
The value of the interdot tunnel coupling $t_{0}$ is obtained by
fitting the experimental data, and thus there may exist some uncertainty
$\alpha_{t_{0}}$ for $t_{0}$ extraction. We regard $\alpha_{t_{0}}$
as a systematic error, that is $\alpha_{t_{0}}$ is a fixed constant
value for a specific two-qubit system, but the fixed constant $\alpha_{t_{0}}$
can vary for different two-qubit systems. Therefore, a more realistic
gate infidelity should be defined as 
\begin{equation}
\mathcal{I}\equiv1-\dfrac{1}{16}\left|{\rm Tr}\left[U_{T}^{\dagger}U_{4\times4}(t_{f})\right]\right|^{2},\label{eq:qd_infidelity-1}
\end{equation}
where $U_{4\times4}(t_{f})$ is the realistic propagator in the subspace
spanned by the two-qubit computational basis states, projected from
the realistic propagator $U(t_{f})=\mathcal{T}_{+}\exp[-(i/\hbar)\int_{0}^{t_{f}}\mathcal{H}(t')dt']$
at $t_{f}$. In general, noise is stochastic, and thus we denote the
ensemble average of gate infidelity $\mathcal{I}$ over the different
noise realizations as $\left\langle \mathcal{I}\right\rangle $.

To characterize the electrical noise $\beta_{U-\epsilon}(t)$, we
simulate the two-qubit dephasing process, the free induction evolution
of the two-qubit system, as shown in Fig.~\ref{fig_prob_simulation_electrical_noise}(a)
or more precisely in Fig.~S6 of the Supplementary Information of
Ref.~\cite{VeldhorstYangHwangEtAl2015}. There, the probability of
the state $\left|\uparrow,\downarrow\right\rangle $, $P(\left|\uparrow,\downarrow\right\rangle )$
(the spin up fraction of dot2 in the $\left|\textrm{dot2},\textrm{dot1}\right\rangle $
basis), for initial state $\left|\downarrow,\downarrow\right\rangle $
after the operations $(\pi/2)_{X_{2}}\rightarrow$ C-phase$(\tau_{Z})$
$\rightarrow(\pi/2)_{Y_{2}}$ with increasing time $\tau_{Z}$ (gate
operation time of C-phase gate) is measured. It was mentioned in the
caption of Fig.~S6 of Ref.~\cite{VeldhorstYangHwangEtAl2015} that
there exists a phase difference $\phi=\pi/2$ separated by the C-phase$(\tau_{Z})$
gate, for which we simulate by inserting a $(\pi/2)_{Z_{2}}$ rotation
between the $(\pi/2)_{X_{2}}$ rotation and the C-phase$(\tau_{Z})$
gate. Here gates $(\pi/2)_{X_{2}}$, $(\pi/2)_{Y_{2}}$, and $(\pi/2)_{Z_{2}}$
represent $\pi/2$ rotations in the $X$-direction, $Y$-direction,
and $Z$-direction, respectively, for the dot2 qubit. To estimate
the strength of the electrical noise causing the two-qubit dephasing
effect shown in Fig.~S6 of Ref.~\cite{VeldhorstYangHwangEtAl2015},
we assume that all single-qubit rotations (gates) are ideal, and thus
the probability loss in Fig.~S6 of Ref.~\cite{VeldhorstYangHwangEtAl2015}
comes entirely from the C-phase gate suffering from the electrical
noise. We model the electrical noise $\beta_{U-\epsilon}(t)$ in the
isotope-enriched silicon QD system having the same experimentally
measured $1/f^{1.01}$ noise spectrum as that in isotope-enriched
$^{28}{\rm Si}/{\rm SiGe}$ quantum dots \cite{YonedaTakedaOtsukaEtAl2018}
in the frequency range between $\omega/2\pi=10^{-2}{\rm Hz}$ and
$10^{6}{\rm Hz}$. To avoid the divergence of $1/f^{1.01}$ noise
spectrum at very low frequency, we assume the noise spectrum $S(\omega)$
gradually saturates to a constant value for frequency $\omega/2\pi<10^{-2}{\rm Hz}$.
This $1/f^{1.01}$ noise spectrum can be simulated via a superposition
of Ornstein-Uhlenbeck processes \cite{Milotti1995,Gleeson2005}. For
the C-phase gate reported in the experiment of Fig.~S6 of Ref.~\cite{VeldhorstYangHwangEtAl2015},
the effective detuning frequency $\nu_{\uparrow\downarrow}=3.14{\rm MHz}$
corresponds to $U-\epsilon=276.71{\rm GHz}$ that can be tuned by
the electrical voltage. Employing the Hamiltonian $\mathcal{H}(t)$
of Eq.\ (\ref{eq:qd_real_hamiltonian}) with these realistic system
parameters, we simulate the ensemble average probability $\left\langle P(\left|\uparrow,\downarrow\right\rangle )\right\rangle $
with increasing time $\tau_{Z}$ for different values of average standard
deviation $\sigma_{U-\epsilon}$ of the electrical noise, each using
a thousand of $\beta_{U-\epsilon}(t)$ noise realizations. We observe
that when $\sigma_{U-\epsilon}$ is chosen to be $2.4{\rm GHz}$,
the corresponding two-qubit coherence (dephasing) time $T_{2,CZ}^{\star}=8.57{\rm \mu s}$
is obtained by fitting the ensemble average probability $\left\langle P(\left|\uparrow,\downarrow\right\rangle )\right\rangle $
with increasing time $\tau_{Z}$ to the formula $\frac{1}{2}+\frac{1}{2}\cos(2\pi\cdot f_{2,CZ}\cdot\tau_{Z})\cdot\exp[-(\tau_{Z}/T_{2,CZ}^{\star})^{a}]$
\cite{YonedaTakedaOtsukaEtAl2018}, where we choose $f_{2,CZ}=\nu_{\uparrow\downarrow}=3.14{\rm MHz}$
and $a=1.9$ for the best fitting result. The simulation data points
(in blue) and the best fitting curve (in red) of $\left\langle P(\left|\uparrow,\downarrow\right\rangle )\right\rangle $
are shown in Fig.\ \ref{fig_prob_simulation_electrical_noise}(a),
and the corresponding noise spectrum $S(\omega)$ and typical noise
realizations $\beta_{U-\epsilon}(t)$ are shown in Figs.\ \ref{fig_prob_simulation_electrical_noise}(b)
and \ref{fig_prob_simulation_electrical_noise}(c), respectively.
This result is very close to the experimentally measured $T_{2,CZ}^{\star}=8.3{\rm \mu s}$
\cite{VeldhorstYangHwangEtAl2015}. Therefore, we use the electrical
noise $\beta_{U-\epsilon}(t)$ with noise spectrum $1/f^{1.01}$ and
average standard deviation $\sigma_{U-\epsilon}=2.4{\rm GHz}$ and
choose $U-\epsilon=276.71{\rm GHz}$ for the following quantum gate
simulations.

\begin{figure}
\includegraphics[scale=0.6]{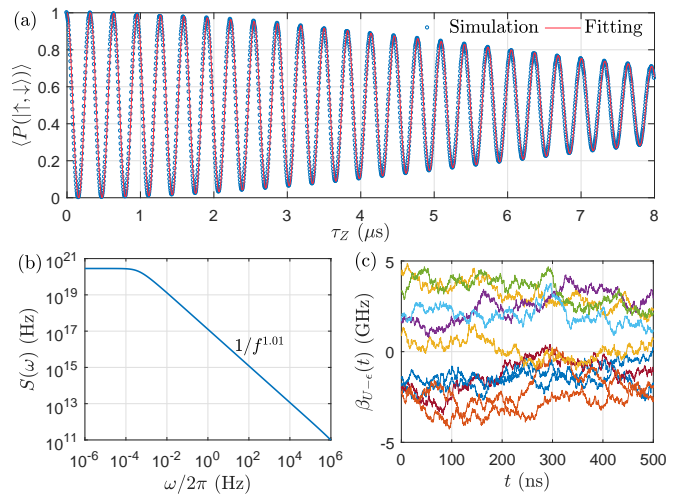}
\protect\protect\protect\caption{Characterization of the electrical noise $\beta_{U-\epsilon}(t)$.
(a) The ensemble average probability $\left\langle P(\left|\uparrow,\downarrow\right\rangle )\right\rangle $
suffering the electrical noise $\beta_{U-\epsilon}(t)$ with the standard
deviation $\sigma_{U-\epsilon}=2.4{\rm GHz}$ is simulated in the
blue circles and fitted in the red line. (b) The corresponding spectrum
$S(\omega)$ of the electrical noise $\beta_{U-\epsilon}(t)$ with
$1/f^{1.01}$ property from $\omega/2\pi=10^{-2}$ to $10^{6}$. (c)
Ten realizations of the corresponding electrical noise $\beta_{U-\epsilon}(t)$.}

\label{fig_prob_simulation_electrical_noise} 
\end{figure}

To suppress the $1/f^{1.01}$ electrical noise we characterize above,
we employ the robust control method \cite{HuangGoan2017} to minimize
the total cost function 
\begin{equation}
\mathcal{K}[\Omega_{X}(t),\Omega_{Y}(t)]=J_{1}+\left\langle J_{2,U-\epsilon}\right\rangle +\xi\,\mathcal{F}\label{eq:total_cost_function}
\end{equation}
via searching the optimal control parameter sets $\left\{ a_{1},a_{2},\cdots,a_{{\rm k_{max}}}\right\} $
and $\left\{ b_{1},b_{2},\cdots,b_{{\rm k_{max}}}\right\} $ in the
respective control pulses $\Omega_{X}(t)=\sum_{k=1}^{k_{{\rm max}}}a_{k}\sin^{3}(\omega_{X,k}\, t)$
and $\Omega_{Y}(t)=\sum_{k=1}^{k_{{\rm max}}}b_{k}\sin^{3}(\omega_{Y,k}\, t)$
where we choose $\omega_{X,k}=(2k-1)\pi/t_{f}$ and $\omega_{Y,k}=(2k)\pi/t_{f}$,
and choose $k_{{\rm max}}=11$ for CNOT gates and $k_{{\rm max}}=8$
for single-qubit gates. The function form of $\sin^{3}(\omega_{X/Y,k}\, t)$
in the control pulses $\Omega_{X}(t)$ and $\Omega_{Y}(t)$ is chosen
to make pulse strengths and pulse slopes vanish at both $t=0$ and
$t=t_{f}$ for smooth pulse-pulse connection to avoid the extra fidelity
loss from the sudden pulse strength change when connecting to their
previous or subsequent gate operations. In the total cost function
$\mathcal{K}$ in Eq.\ (\ref{eq:total_cost_function}), $J_{1}$
is the ideal gate infidelity, and $\left\langle J_{2,U-\epsilon}\right\rangle =$$\frac{1}{2}\int_{0}^{t_{f}}2\pi dt_{1}\int_{0}^{t_{1}}2\pi dt_{2}C_{U-\epsilon}(t_{1},t_{2})$
$\times{\rm Re}\{{\rm Tr}[(R_{U-\epsilon}(t_{1})R_{U-\epsilon}(t_{2}))_{4\times4}]\}$$-\frac{1}{16}\int_{0}^{t_{f}}2\pi dt_{1}\int_{0}^{t_{f}}2\pi dt_{2}$
$C_{U-\epsilon}(t_{1},t_{2}){\rm Tr}[R_{U-\epsilon,4\times4}(t_{1})]{\rm Tr}[R_{U-\epsilon,4\times4}(t_{2})]$
is the lowest order contribution from the electrical noise $\beta_{U-\epsilon}(t)$
in the ensemble average infidelity $\left\langle \mathcal{I}\right\rangle $,
where $C_{U-\epsilon}(t_{1},t_{2})=\left\langle \beta_{U-\epsilon}(t_{1})\beta_{U-\epsilon}(t_{2})\right\rangle $
is the correlation function of the electrical noise and can be obtained
from the noise spectrum $S(\omega)$ in Fig.\ \ref{fig_prob_simulation_electrical_noise}(b)
via the Wiener-Khinchin theorem, i.e., $C_{U-\epsilon}(t_{1},t_{2})=C_{U-\epsilon}(t_{1}-t_{2})=\frac{1}{2\pi}\int_{-\infty}^{\infty}S(\omega)\, e^{i\omega(t_{1}-t_{2})}d\omega$
\cite{MillerChilders2012a}. The operators $[R_{U-\epsilon}(t_{1})R_{U-\epsilon}(t_{2})]_{4\times4}$
and $R_{U-\epsilon,4\times4}(t)$ are $R_{U-\epsilon}(t_{1})R_{U-\epsilon}(t_{2})$
and $R_{U-\epsilon}(t)$ projected onto the subspace spanned by the
computational basis states, respectively. Here $R_{U-\epsilon}(t)\equiv U_{I}^{\dagger}(t)H_{U-\epsilon}U_{I}(t)$,
$U_{I}(t)$ is the ideal propagator obtained by the ideal Hamiltonian
$\mathcal{H}_{I}(t)$ in Eq.\ (\ref{eq:qd_ideal_hamiltonian}), and
$H_{U-\epsilon}$ is a $5\times5$ matrix of the electrical noise
Hamiltonian with all matrix elements being zeros except one element
with value being one in the location of $U-\epsilon(t)$ in Eq.\ (\ref{eq:qd_ideal_hamiltonian}),
i.e., $H_{U-\epsilon}(5,5)=1$. The quantity $\mathcal{F}$ in the
last term of Eq.\ (\ref{eq:total_cost_function}) defined as $\mathcal{F}\equiv\int_{0}^{t_{f}}\left|\Omega_{X}(t)\right|^{2}dt+\int_{0}^{t_{f}}\left|\Omega_{Y}(t)\right|^{2}dt$$+\left|\int_{0}^{t_{f}}\left|\Omega_{X}(t)\right|^{2}dt-\int_{0}^{t_{f}}\left|\Omega_{Y}(t)\right|^{2}dt\right|$
is the fluence (a measure of the field energy) \cite{KosutGraceBrif2013},
which is used to restrain or minimize the strengths of ac magnetic
field control pulses $\Omega_{X}(t)$ and $\Omega_{Y}(t)$. The factor
$\xi$ also in the same last term of Eq.\ (\ref{eq:total_cost_function})
determines the contribution ratio of the fluence $\mathcal{F}$ to
the ensemble average infidelity in the total cost function. If $\xi$
is too small, $\mathcal{F}$ does not work. However, if $\xi$ is
too large, then $\mathcal{F}$ dominates the contribution in the cost
function and thus $\left\langle J_{2,U-\epsilon}\right\rangle $ may
not be effectively suppressed. In our simulations, we find $\xi=10^{-6}$
works effectively.

To meet the constraint of maximum pulse strength $1{\rm mT}$, we
suitably choose the gate operation time $t_{f}=500{\rm ns}$ for CNOT
gates and $t_{f}=200{\rm ns}$ and $250{\rm ns}$ for single-qubit
$I_{2}\otimes X_{1}$ gate (Identity gate for dot2 qubit and $X$
gate for dot1 qubit) and $H_{2}\otimes I_{1}$ gate (Hadamard gate
for dot2 qubit and Identity gate for dot1 qubit), respectively. After
running the optimization procedure, we obtain the optimal gate pulses
that can suppress the electrical noise while keeping the maximum strength
of the optimal ac magnetic field control pulses smaller than ${\rm 1mT}$.
However, due to the finite bandwidth of waveform generators the actual
output pulses $\Omega_{X}^{{\rm filt}}(t)$ and $\Omega_{Y}^{{\rm filt}}(t)$
on the qubits will be distorted as compared to the input optimal pulses
$\Omega_{X}(t)$ and $\Omega_{Y}(t)$. The filtering effects of the
waveform generators can be modeled via the transfer function $\Omega^{{\rm filt}}(t)=\frac{1}{2\pi}\int_{-\infty}^{+\infty}d\omega e^{i\omega t}F(\omega)\Omega(\omega)$,
where $\Omega(\omega)=\int_{-\infty}^{+\infty}dt'e^{-i\omega t'}\Omega(t')$
is the input optimal pulse in the frequency domain, and $F(\omega)=\exp(-\omega^{2}/\omega_{c}^{2})$
is the response function of the filter with $\omega_{c}$ being the
cutoff frequency \cite{TheisMotzoiWilhelm2016,MotzoiGambettaMerkelEtAl2011}.
We use the value of $\omega_{c}/2\pi=425.4{\rm MHz}$ (approximation
for Tektronix AWG5014 \cite{TheisMotzoiWilhelm2016,MotzoiGambettaMerkelEtAl2011})
for simulating the filtering effects on the quantum gates demonstrated
here. The pulse distortion due to the filtering effects will degrade
the gate fidelity from the expected value. Thus, we perform an extra
fine-tuning optimization with the same cost function $\mathcal{K}$
in Eq.\ (\ref{eq:total_cost_function}) but replacing the unfiltered
pulses $\Omega_{X}(t)$ and $\Omega_{Y}(t)$ with the filtered pulses
$\Omega_{X}^{{\rm filt}}(t)$ and $\Omega_{Y}^{{\rm filt}}(t)$ to
obtain the fine-tuned optimal gate pulses to recover the fidelity
loss. For computation efficiency \cite{supplementary}, we calculate
the total cost function $\mathcal{K}$ in Eq.\ (\ref{eq:total_cost_function})
to obtain the optimal control pulses using an effective Hamiltonian
with the rotating-wave approximation and the second-order approximation
after the Schrieffer-Wolff transformation \cite{MeunierCaladoVandersypen2011},
while we use the full realistic Hamiltonian $\mathcal{H}(t)$ of Eq.\ (\ref{eq:qd_real_hamiltonian})
without these approximations to simulate the ensemble average infidelity
$\left\langle \mathcal{I}\right\rangle $ with an ensemble of one
thousand noise realizations for demonstrating the gate performance.

\begin{figure}
\includegraphics[scale=0.6]{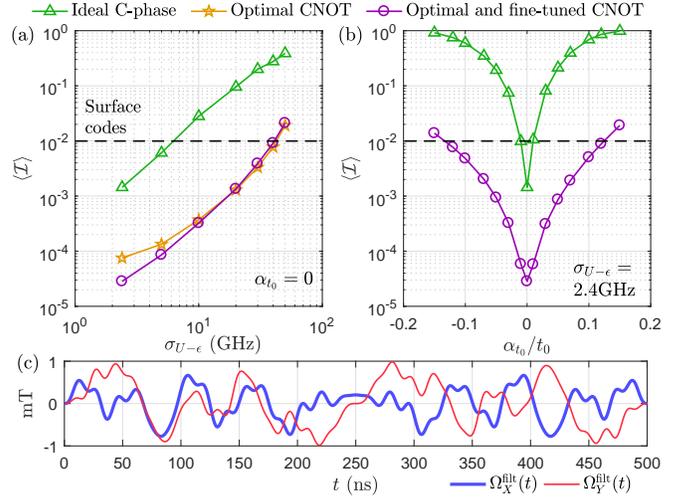}
\protect\protect\protect\caption{The robust performance (a) against the electrical noise $\beta_{U-\epsilon}(t)$
with $1/f^{1.01}$ noise spectrum and (b) against the uncertainty
$\alpha_{t_{0}}$ in $t_{0}$ under $\sigma_{U-\epsilon}=2.4{\rm GHz}$
for the ideal C-phase gate in the green-triangle line, the optimal
CNOT gate without fine-tuning optimization in the orange-pentagram
line, and the optimal CNOT gate with fine-tuning optimization in the
purple-circle line. (c) The actual output pulses with the filtering
effects, $\Omega_{X}^{{\rm filt}}(t)$ in bold-blue line and $\Omega_{Y}^{{\rm filt}}(t)$
in thin-red line, for the optimal and fine-tuned CNOT gate.}

\label{fig_cnot_gate_performance} 
\end{figure}

The robust performance against the electrical noise $\beta_{U-\epsilon}(t)$
with $1/f^{1.01}$ noise spectrum is shown in Fig.\ \ref{fig_cnot_gate_performance}(a).
The optimal CNOT gate with the fine-tuning optimization in the purple-circle
line can recover the degradation in the ensemble average infidelity
$\left\langle \mathcal{I}\right\rangle $ resulting from the filtering
effects (in the orange-pentagram line) by about half-order of magnitude
for smaller $\sigma_{U-\epsilon}$. For larger $\sigma_{U-\epsilon}$,
the contribution of infidelity increase due to the filtering effect
is much smaller than that due to the electrical noise so that no considerable
improvement is observed. However, the fine-tuned optimal CNOT gate
can improve the ensemble average infidelity $\left\langle \mathcal{I}\right\rangle $
over the ideal C-phase gate in the green-triangle line by near two
orders of magnitude at $\sigma_{U-\epsilon}=2.4{\rm GHz}$, and for
gate error (infidelity) less than the error threshold of surface codes
$\left\langle \mathcal{I}\right\rangle \apprle10^{-2}$ \cite{FowlerMariantoniMartinisEtAl2012},
the fine-tuned optimal CNOT gate can be robust against the noise strength
to $\sigma_{U-\epsilon}\cong40{\rm GHz}$ while the ideal C-phase
gate can be robust only to $\sigma_{U-\epsilon}\cong6{\rm GHz}$.
The uncertainty error $\alpha_{t_{0}}$ in tunnel coupling $t_{0}$
appears in the same location as $\beta_{U-\epsilon}(t)$ in the effective
Hamiltonian after the Schrieffer-Wolff transformation \cite{MeunierCaladoVandersypen2011},
and thus the constructed optimal gate pulses robust against the electrical
noise will be also robust against the uncertainty error $\alpha_{t_{0}}$
in $t_{0}$. This can be seen in Fig.\ \ref{fig_cnot_gate_performance}(b)
that our fine-tuned optimal CNOT gate at the electrical noise $\sigma_{U-\epsilon}=2.4{\rm GHz}$
can be robust against uncertainty $\alpha_{t_{0}}$ to about $12\%$
of $t_{0}$ (in the purple-circle line), while the ideal C-phase gate
can be robust against $\alpha_{t_{0}}$ to only about $1\%$ of $t_{0}$
(in the green-triangle line) for $\left\langle \mathcal{I}\right\rangle \apprle10^{-2}$.
The actual output pulses on the qubits with the filtering effects
of the fine-tuned optimal CNOT gate in Figs.\ \ref{fig_cnot_gate_performance}(a)
and \ref{fig_cnot_gate_performance}(b) are shown in Fig.\ \ref{fig_cnot_gate_performance}(c),
and the maximum strengths of $\left|\Omega_{X}^{{\rm filt}}(t)\right|$
and $\left|\Omega_{Y}^{{\rm filt}}(t)\right|$ within the gate operation
time $t_{f}=500{\rm ns}$ are all smaller than $1{\rm mT}$. Under
the same experimental constraints, control framework, and voltage
setting as the two-qubit CNOT gate in Fig.\ \ref{fig_cnot_gate_performance},
high-fidelity and robust single-qubit gates can also be realized.
For example, we find that our optimal $I_{2}\otimes X_{1}$ and $H_{2}\otimes I_{1}$
gates with the fine-tuning optimization can be robust against the
electrical noise to $\sigma_{U-\epsilon}\cong{\rm 50GHz}$ for $\left\langle \mathcal{I}\right\rangle \apprle10^{-2}$,
and at $\sigma_{U-\epsilon}=2.4{\rm GHz}$ the gate infidelity $\left\langle \mathcal{I}\right\rangle \cong2\times10^{-5}$.
We also investigate the performance of the CNOT gate for $1/f^{\alpha}$
electrical noise spectra with $0.7\leq\alpha<1.01$, which have larger
high-frequency contributions than the $1/f^{1.01}$ noise spectrum.
As $\alpha$ decreases from $1.01$ to $0.7$, the $\left\langle \mathcal{I}\right\rangle $
of the fine-tuned optimal CNOT gate constructed via the same robust
control method \cite{HuangGoan2017} gradually increases from $2.84\times10^{-5}$
to $2.5\times10^{-4}$, but still having improvement from two-order
to one-order over the ideal C-phase gate.

In summary, we have constructed a high-fidelity CNOT gate and single-qubit
gates robust against the time-varying electrical noise $\beta_{U-\epsilon}(t)$
with the experimentally measured $1/f^{1.01}$ noise spectrum and
against the system parameter uncertainty $\alpha_{t_{0}}$ in $t_{0}$.
In our proposed control framework, the detuning $\epsilon$ is kept
constant for all single- and two-qubit gate operations to avoid possible
extra errors coming from tuning $\epsilon$ up and down for a sequence
of gate operations. We control only two experimentally realizable
ac magnetic fields with pulse strengths satisfying the constraint
of the device. Our scheme that can also recover the fidelity loss
from the filtering effects will provide an essential step toward large-scale
FTQC for quantum-dot spin qubits in isotopically enriched silicon. 
\begin{acknowledgments}
We acknowledge support from the the Ministry of Science and Technology
of Taiwan under Grant No.~MOST 106-2112-M-002-013-MY3, from the National
Taiwan University under Grant No.~NTU-CCP-106R891703, and from the
thematic group program of the National Center for Theoretical Sciences,
Taiwan. C.H.Y. and A.S.D. acknowledge support from the Australian
Research Council (CE11E0001017) and the US Army Research Office (W911NF-17-1-0198). 
\end{acknowledgments}

\bibliographystyle{apsrev4-1}
\bibliography{citation_robust_gates}

\clearpage

\onecolumngrid

\section*{Supplementary material for ``High-fidelity and robust two-qubit
gates for quantum-dot spin qubits in silicon''}

\global\long\def\theequation{S-\arabic{equation}}

\section{Gate infidelity contribution form single-qubit dephasing noise }

We evaluate the gate infidelity contributed from the residual single-qubit
decoherence. The dephasing noise spectrum of the ${\rm ^{31}P}$ electron
spin qubit in the isotopically enriched silicon was measured to be
$S(\omega)\propto1/f^{2.5}+c$ in the frequency range of $10^{3}-10^{5}{\rm Hz}$,
where $c$ is a constant and the plausible sources of the low-frequency
noise could be thermal noise and magnetic noise due to the instability
of the external magnetic field {[}1{]}. Here we use the same form
of noise spectrum for our quantum-dot spin qubits as they both are
in the same isotopically enriched ${\rm ^{28}Si}$ substrate. For
the quantum-dot electron spin qubit in the ${\rm GaAs}$ substrate,
the dephasing noise spectrum $\sim1/f^{2.6}$ in the frequency range
of $10^{1}-10^{5}{\rm Hz}$ has also been observed {[}2{]}. To extract
the correct noise strength for our target silicon quantum-dot system
{[}3{]}, we use an ensemble of the dephasing noise realizations for
different average noise standard deviations to simulate the decay
of the single-qubit Ramsey fringe oscillations in Ref.~{[}4{]} with
the formula $\frac{1}{2}+\frac{1}{2}\exp(-(t/T_{2}^{\star})^{n})$
to find the best fitting result for the $T_{2}^{\star}=120\mu s$
and $n=2$ {[}1{]}. The contribution to the ensemble average infidelity
of our fine-tuned optimal CNOT gate from two independent dephasing
noises on the two quantum dots with the obtained fitted noise strengths
is $\sim3.43\times10^{-6}$, which is around one order of magnitude
smaller than $\left\langle \mathcal{I}\right\rangle \cong2.84\times10^{-5}$
obtained from the simulation for the electrical noise of $\sigma_{U-\epsilon}=2.4{\rm GHz}$.
Therefore, the residual single-qubit dephasing noise in our target
system does not affect the performance of the fine-tuned optimal CNOT
gate demonstrated above. Besides, the contribution of the dephasing
noise to the ensemble average infidelity of the ideal C-phase gate
constructed without considering any noise effect is $\sim2.19\times10^{-5}$,
one order of magnitude larger than that of our CNOT gate. This shows
that our optimal pulses have the ability to also suppress the dephasing
noise even though we do not include the dephasing noise contribution
into our total cost function for optimization.

\section{Effective Hamiltonian for optimization}

Here, we explain in more detail how we obtain the effective Hamiltonian
for optimization calculations. The values of the system parameters
used in our Hamiltonian vary quite a lot: $\overline{E}_{Z}=39.16{\rm GHz}$,
$\delta E_{Z}=-40{\rm MHz}$, $t_{0}=900{\rm MHz}$, and $U-\epsilon=276.71{\rm GHz}$.
The largest value in these parameters is over 6900 times greater than
the smallest one. Thus to obtain the exact dynamics, one needs to
choose much smaller time-step for computation i.e., requires very
long computation time. For computation efficiency in our numerical
optimization, we apply two approximations to the Hamiltonian: the
first one is to use the Schrieffer-Wolff (SW) transformation {[}5{]}
and keep terms up to the second order, and the other one is to apply
the rotating-wave approximation (RWA). Once we obtain the optimal
control pulses to suppress the gate error from the noise, we will
input the control pulses to the full Hamiltonian without using these
approximations to calculate the gate infidelity.

Using the SW transformation, we transform the ideal Hamiltonian $\mathcal{H}_{I}(t)$
to 
\begin{equation}
\widetilde{\mathcal{H}}_{I}^{{\rm SW}}(t)=e^{S}\mathcal{H}_{I}(t)e^{-S},\label{eq:ideal_hamiltonian_sw}
\end{equation}
where 
\begin{align}
S & =\left(\begin{array}{ccccc}
0 & 0 & 0 & 0 & 0\\
0 & 0 & 0 & 0 & -\gamma(-\delta E_{Z})\\
0 & 0 & 0 & 0 & \gamma(\delta E_{Z})\\
0 & 0 & 0 & 0 & 0\\
0 & \gamma(-\delta E_{Z}) & -\gamma(\delta E_{Z}) & 0 & 0
\end{array}\right),
\end{align}
and 
\begin{equation}
\gamma(\delta E_{Z})=\dfrac{t_{0}}{U-\epsilon+\delta E_{Z}/2}.
\end{equation}
For $(U-\epsilon)\gg t_{0}$ and $(U-\epsilon)\gg\left|\delta E_{Z}\right|$,
we can expand $\mathcal{\widetilde{H}}_{I}^{{\rm SW}}(t)$ in Eq.\ (\ref{eq:ideal_hamiltonian_sw})
to the second order of $S$ and omit the terms including $O[\gamma^{2}(\delta E_{Z})]$
or $[\gamma(-\delta E_{Z})-\gamma(\delta E_{Z})]$ to obtain the Hamiltonian
\begin{equation}
\mathcal{\widetilde{H}}_{I}^{{\rm SWA}}(t)=\left(\begin{array}{cc}
\mathcal{\widetilde{H}}_{I,4\times4}^{{\rm SWA}}(t) & \begin{array}{c}
0\\
0\\
0\\
0
\end{array}\\
\begin{array}{cccc}
0 & 0 & 0 & 0\end{array} & \mathcal{\widetilde{H}}_{I}^{{\rm SWA}}(5,5)
\end{array}\right),
\end{equation}
where 
\begin{equation}
\mathcal{\widetilde{H}}_{I,4\times4}^{{\rm SWA}}(t)=h\left(\begin{array}{cccc}
\overline{E}_{Z} & \frac{1}{2}E_{X}(t) & \frac{1}{2}E_{X}(t) & 0\\
\frac{1}{2}E_{X}(t) & \frac{1}{2}\delta E_{Z}-J_{m} & \frac{1}{2}\left(J_{p}+J_{m}\right) & \frac{1}{2}E_{X}(t)\\
\frac{1}{2}E_{X}(t) & \frac{1}{2}\left(J_{p}+J_{m}\right) & -\frac{1}{2}\delta E_{Z}-J_{m} & \frac{1}{2}E_{X}(t)\\
0 & \frac{1}{2}E_{X}(t) & \frac{1}{2}E_{X}(t) & -\overline{E}_{Z}
\end{array}\right),\label{eq:ideal_hamiltonian_swa_4x4}
\end{equation}
$\mathcal{\widetilde{H}}_{I}^{{\rm SWA}}(5,5)=U-\epsilon+J_{p}+J_{m}$,
and 
\begin{align}
J_{p} & \equiv\dfrac{t_{0}^{2}}{U-\epsilon+\delta E_{Z}/2},\\
J_{m} & \equiv\dfrac{t_{0}^{2}}{U-\epsilon-\delta E_{Z}/2}.
\end{align}
The superscripts SWA denote the Hamiltonian with the above approximation
after the Schrieffer-Wolff transformation. The elements of the Hamiltonian
$\mathcal{\widetilde{H}}_{I}^{{\rm SWA}}(t)$ in the subspace spanned
by the computational basis states $\left\{ \left|\uparrow,\uparrow\right\rangle ,\left|\uparrow,\downarrow\right\rangle ,\left|\downarrow,\uparrow\right\rangle ,\left|\downarrow,\downarrow\right\rangle \right\} $
and in the subspace of $\left|0,2\right\rangle $ are decoupled. Therefore,
we can simulate the dynamics of the system in the above two subspaces
separately.

Since the strengths of the control pulses $\left|\Omega_{X}(t)\right|$
and $\left|\Omega_{Y}(t)\right|$ are constrained to be smaller than
$1{\rm mT}$, the maximum value of $\frac{1}{2}\left|E_{X}(t)\right|$
is at most $\sim28{\rm MHz}$, which is over 1000 times smaller than
$\overline{E}_{Z}=39.16{\rm GHz}$. Thus, we can apply the RWA to
the Hamiltonian. Transforming $\mathcal{\widetilde{H}}_{I,4\times4}^{{\rm SWA}}(t)$
to the rotating frame (RF), we obtain the Hamiltonian in the computational
state basis as 
\begin{equation}
\widetilde{\mathcal{H}}_{I,4\times4}^{{\rm SWA,RF}}(t)=U_{0}^{\dagger}(t)\mathcal{\widetilde{H}}_{I,4\times4}^{{\rm SWA}}(t)U_{0}(t)-i\hbar U_{0}^{\dagger}(t)\dot{U_{0}}(t),\label{eq:ideal_hamiltonian_swa_rf_4x4}
\end{equation}
where 
\begin{equation}
U_{0}(t)=\left(\begin{array}{cccc}
\exp(-i\overline{E}_{Z}2\pi t) & 0 & 0 & 0\\
0 & 1 & 0 & 0\\
0 & 0 & 1 & 0\\
0 & 0 & 0 & \exp(+i\overline{E}_{Z}2\pi t)
\end{array}\right).\label{eq:u0_rf_transformation}
\end{equation}
Then, by making the RWA, Eq.~(\ref{eq:ideal_hamiltonian_swa_rf_4x4})
becomes 
\begin{align}
 & \mathcal{\widetilde{H}}_{I,4\times4}^{{\rm SWA,RWA}}(t)\nonumber \\
 & =h\left(\begin{array}{cccc}
0 & \frac{1}{4}\overline{\Omega}_{X}(t)-i\frac{1}{4}\overline{\Omega}_{Y}(t) & \frac{1}{4}\overline{\Omega}_{X}(t)-i\frac{1}{4}\overline{\Omega}_{Y}(t) & 0\\
\frac{1}{4}\overline{\Omega}_{X}(t)+i\frac{1}{4}\overline{\Omega}_{Y}(t) & \frac{1}{2}\delta E_{Z}-J_{m} & \frac{1}{2}\left(J_{p}+J_{m}\right) & \frac{1}{4}\overline{\Omega}_{X}(t)-i\frac{1}{4}\overline{\Omega}_{Y}(t)\\
\frac{1}{4}\overline{\Omega}_{X}(t)+i\frac{1}{4}\overline{\Omega}_{Y}(t) & \frac{1}{2}\left(J_{p}+J_{m}\right) & -\frac{1}{2}\delta E_{Z}-J_{m} & \frac{1}{4}\overline{\Omega}_{X}(t)-i\frac{1}{4}\overline{\Omega}_{Y}(t)\\
0 & \frac{1}{4}\overline{\Omega}_{X}(t)+i\frac{1}{4}\overline{\Omega}_{Y}(t) & \frac{1}{4}\overline{\Omega}_{X}(t)+i\frac{1}{4}\overline{\Omega}_{Y}(t) & 0
\end{array}\right),\label{eq:ideal_hamiltonian_swa_rwa_4x4}
\end{align}
where $\overline{\Omega}_{X}(t)\equiv\dfrac{g\mu_{B}}{h}\Omega_{X}(t)$
and $\overline{\Omega}_{Y}(t)\equiv\dfrac{g\mu_{B}}{h}\Omega_{Y}(t)$.
After the above two approximations (SWA and RWA), the parameters in
the Hamiltonian $\mathcal{\widetilde{H}}_{I,4\times4}^{{\rm SWA,RWA}}(t)$
range only from $\sim2.9{\rm MHz}$ to $40{\rm MHz}$, and thus we
can save a lot of computation time to obtain the propagator $\widetilde{U}_{I,4\times4}^{{\rm SWA,RWA}}(t)$
of the Hamiltonian $\mathcal{\widetilde{H}}_{I,4\times4}^{{\rm SWA,RWA}}(t)$
by the Schr$\ddot{{\rm o}}$dinger equation. Then transforming this
propagator $\widetilde{U}_{I,4\times4}^{{\rm SWA,RWA}}(t)$ from the
rotating frame back to the frame transformed by the SW transformation
and combining it with the propagator in the subspace $\left|0,2\right\rangle $
in the same frame, we obtain the propagator in the full space 
\begin{equation}
\widetilde{U}_{I}^{{\rm SWA}}(t)=\left(\begin{array}{cc}
\widetilde{U}_{I,4\times4}^{{\rm SWA}}(t) & \begin{array}{c}
0\\
0\\
0\\
0
\end{array}\\
\begin{array}{cccc}
0 & 0 & 0 & 0\end{array} & \exp\left(-i\{U-\epsilon+J_{p}+J_{m}\}2\pi t\right)
\end{array}\right).
\end{equation}
Finally, the ideal system propagator in the original frame, $U_{I}(t)$,
is obtained via the transformation 
\begin{equation}
U_{I}(t)\cong e^{-S}\widetilde{U}_{I}^{{\rm SWA}}(t)e^{+S},
\end{equation}
where we expand $e^{-S}$ and $e^{+S}$ to the second order of $S$.
Finally, we substitute the propagator $U_{I}(t)$ into the total cost
function $\mathcal{K}$ of Eq.~(6) of the main text for optimization
to find the control pulses. However, to calculate the performance
of our gates, we apply the obtained optimal control pulses to the
full realistic Hamiltonian $\mathcal{H}(t)$ of Eq.~(3) of the main
text without these approximations to simulate the ensemble average
infidelity $\left\langle \mathcal{I}\right\rangle $ with an ensemble
of one thousand noise realizations. 

\vspace{3mm}
\begin{lyxlist} {00.00}
\item [{{[}1{]}}] J. T. Muhonen, J. P. Dehollain, A. Laucht, F. E. Hudson,
R. Kalra, T. Sekiguchi, K. M. Itoh, D. N. Jamieson, J. C. McCallum,
A. S. Dzurak, and A. Morello, Nat Nano 9, 986 (2014).
\item [{{[}2{]}}] J. Medford, {L. }Cywi\ifmmode \acute{n}\else \'{n}\fi{}ski,
C. Barthel, C. M. Marcus, M. P. Hanson, and A. C. Gossard, Phys. Rev.
Lett. 108, 086802 (2012).
\item [{{[}3{]}}] M. Veldhorst, C. H. Yang, J. C. C. Hwang, W. Huang, J.
P. Dehollain, J. T. Muhonen, S. Simmons, A. Laucht, F. E. Hudson,
K. M. Itoh, A. Morello, and A. S. Dzurak, Nature 526, 410 (2015).
\item [{{[}4{]}}] M. Veldhorst, J. C. C. Hwang, C. H. Yang, A. W. Leenstra,
B. de Ronde, J. P. Dehollain, J. T. Muhonen, F. E. Hudson, K. M. Itoh,
A. Morello, and A. S. Dzurak, Nat. Nanotech. 9, 981 (2014).
\item [{{[}5{]}}] T. Meunier, V. E. Calado, and L. M. K. Vandersypen, Phys.
Rev. B 83, 121403 (2011). \end{lyxlist}

\end{document}